\documentclass[extra,onecolumn]{gji}
\usepackage{graphicx,color}
\usepackage{amsmath}
\title[Modelling the Southern African Geomagnetic Field]
  {Regional Modelling of the Southern African Geomagnetic Field using Harmonic Splines}
\author[A. Geese \textit{et. al.}]
  {A. Geese$^{1,2}$, M. Hayn$^3$, M. Mandea$^{1,4}$, V. Lesur$^1$\\
  $^1$ Helmholtz-Zentrum Potsdam, Deutsches GeoForschungsZentrum - GFZ, Telegrafenberg, 14473 Potsdam, Germany\\
  $^2$ Institut f\"ur Geophysik und extraterrestrische Physik, TU Braunschweig, Mendelssohnstr. 3, 38106 Braunschweig, Germany\\
  $^3$ Universit\"at Potsdam, Institut f\"ur Mathematik, Am Neuen Palais 10, 14469 Potsdam, Germany\\
  $^4$ now at Universit\'{e} Paris Diderot - Institut de Physique du Globe de Paris, G\'{e}ophysique spatiale et plan\'{e}taire - \\ B\^{a}timent Lamarck,  Case 7011, 5 rue Thomas Mann, 75205 Paris Cedex 13, France
  }
\date{Received 2009 November 2; in original form 2009 July 30}
\pagerange{\pageref{firstpage}--\pageref{lastpage}}
\volume{200}
\pubyear{2009}

\begin{document}
\label{firstpage}
\maketitle
\begin{summary}
Over the southern African region the geomagnetic field is weak and changes rapidly. For this area series of geomagnetic field measurements exist since the 1950s. We take advantage of the existing repeat station surveys and observatory annual means, and clean these data sets by eliminating jumps and minimising external field contributions in the original time series. This unique data set allows us to obtain a detailed view of the geomagnetic field behaviour in space and time by computing a regional model. For this, we use a system of representation similar to harmonic splines. Initially, the technique is systematically tested on synthetic data. After systematically testing the method on synthetic data, we derive a model for 1961 to 2001 that gives a detailed view of the fast changes of the geomagnetic field in this region.
\end{summary}
\section{Introduction}
The Earth's magnetic field is mainly due to a geodynamo mechanism in the liquid iron outer core. The lithospheric contribution, due to rocks which acquired information about the magnetic field at the time of their solidification from the molten state, adds to the dominant core magnetic field. In addition, external fields represent a third contribution which is produced by the interaction of the solar wind with the magnetosphere and the ionosphere. Their intensities vary with the solar wind speed and the orientation of the embedded magnetic field. The solar activity modifies current systems in the magnetosphere and ionosphere surrounding the Earth, producing magnetic variations on time scales varying from a second to a solar cycle. Moreover, these highly variable external fields cause secondary, induced fields in oceans and electrically conductive parts of the lithosphere and the upper mantle. We can see that the geomagnetic field has a complicated behaviour, varying in space and time.\\
The total intensity of the geomagnetic field  is over $60.000\mbox{ nT}$ in the polar regions. Over the southern African region and the southern Atlantic Ocean it decreases to about $23.000\mbox{ nT}$. In this area, the field is not only weak, but is also changing in intensity and direction very rapidly \citep{Kotze2003, MandeaEtAl2007, LesurEtAl2008}. Considering the long time series from Cape Town and subsequently Hermanus Magnetic Observatories at the southern tip of Africa, the magnetic field strength decreased by about 26\% since the 1920s. This makes the southern African region of particular interest, the aim being to improve the existing models over this region. For this we take advantage of the remarkable set of ground-based measurements available over this part of the continent. This set consists in geomagnetic observatories and repeat station data.\\
The South African data have been used in different modelling approaches on global and regional scales. Let us note that global models such as the IGRF model  \citep{IGRF10} or more complex models (CM4 \citep{CM4}, GRIMM \citep{LesurEtAl2008}, or CHAOS \citep{OlsenEtAl2006}) are built using spherical harmonic expansions. Their spatial resolution depends on the order of the expansion, meaning thousands of coefficients must be estimated to resolve structures in the order of hundreds of kilometers (e.g., an expansion of degree $L=80$ involves 6560 coefficients and resolves wavelengths greater than $500$ km). Any small scale and rapidly varying structures that are present in the southern African region are too small to be described by the available global field models. A regional model should be able to monitor smaller features and to track their fast changes. For this reason we are interested in a better description of the magnetic field on this regional scale, using a different method for regional field modelling.\\
Several approaches to regional modelling of potential fields have been undertaken. We give here a short overview of the most commonly used methods.
\begin{itemize}
\item{Assuming that the area is small enough such that the curvature of the Earth can be neglected, a model can be derived using a surface polynomial. This approach is often applied for regional models, as has even been the case for our region of interest \citep{Kotze2007}. These models are suitable for technical use (e.g. for getting declination maps for navigation), but do not allow for a deeper  understanding of the physical processes which drive the observed spatial and temporal changes. Furthermore, including different data types such as ground-based, airborne or satellite measurements is not possible.}
\item{Spherical cap harmonic analysis (SCHA) was first introduced by \cite{Haines1985}
for modelling MAGSAT satellite data over Canada. The basis functions are only defined on a spherical
cap. Unfortunately, these basis functions are not orthogonal and regularization
becomes difficult: coefficients can be rejected because of statistical insignificance, or based on physical arguments \citep{KorteHolme2003}. \citet{ThebaultEtal2004} have stressed that  exact upward continuation can be incorporated properly only if a 3-D boundary value problem is solved. Nevertheless, SCHA has been used for regional magnetic field modelling (e.g. \citet{NevanlinnaEtAl1988,DeSantisTorta2004,TohEtAl2007}), but the  physical interpretation of these models remains a difficult task.}
\item{During the last few years, several authors \cite[]{HolschneiderEtAl2003, ChambodutEtAl2005} have proposed to use wavelet frames for constructing regional potential field models. Unlike the basis functions, the
elements of a frame are not necessarily linearly independent. As an example of implementation of the wavelet representation for potential fields we note the work of \cite{PanetEtAl2006}. They applied the wavelet method to refine a global, ''low''-resolution gravity field with local, high-resolution data, leading to a combined gravity field in French Polynesia. For the same area, they also obtained a crustal magnetic field model, based only on CHAMP satellite data.}
\end{itemize}
The different modelling approaches summarised above have been introduced because of the general need for high-quality regional models. We consider the advantages and drawbacks of the previous techniques and turn to harmonic splines. This system of representation has already been used for smoothing data on the sphere \citep{Wahba1981} and is suitable for regional modelling. \cite{ShureEtAL1982} introduced this method for modelling the Earth's global magnetic field. Moreover, \cite{Whaler1994} applied harmonic splines to model the lithospheric field over the African continent. Since the 1990's, the harmonic spline technique has not been used extensively. We turn back to it and apply it to model the changing magnetic field over the southern African region.\\
The paper is organised as follows. We first discuss the available data set over the  southern African region and compare it with the long-term CM4 model \citep{CM4}, already indicating some interesting features. The next section deals with the mathematical background for the spline modelling. The results obtained  when this approach is applied to  a synthetic data set computed from the CM4  model are thereafter presented. In the last section, the new harmonic splines model based on real data is shown and discussed. In the conclusion, we list some unsolved problems and propose solutions which may serve to address them.\\

\section{Data}\label{data}
\begin{figure}
\center
\includegraphics[width=8cm]{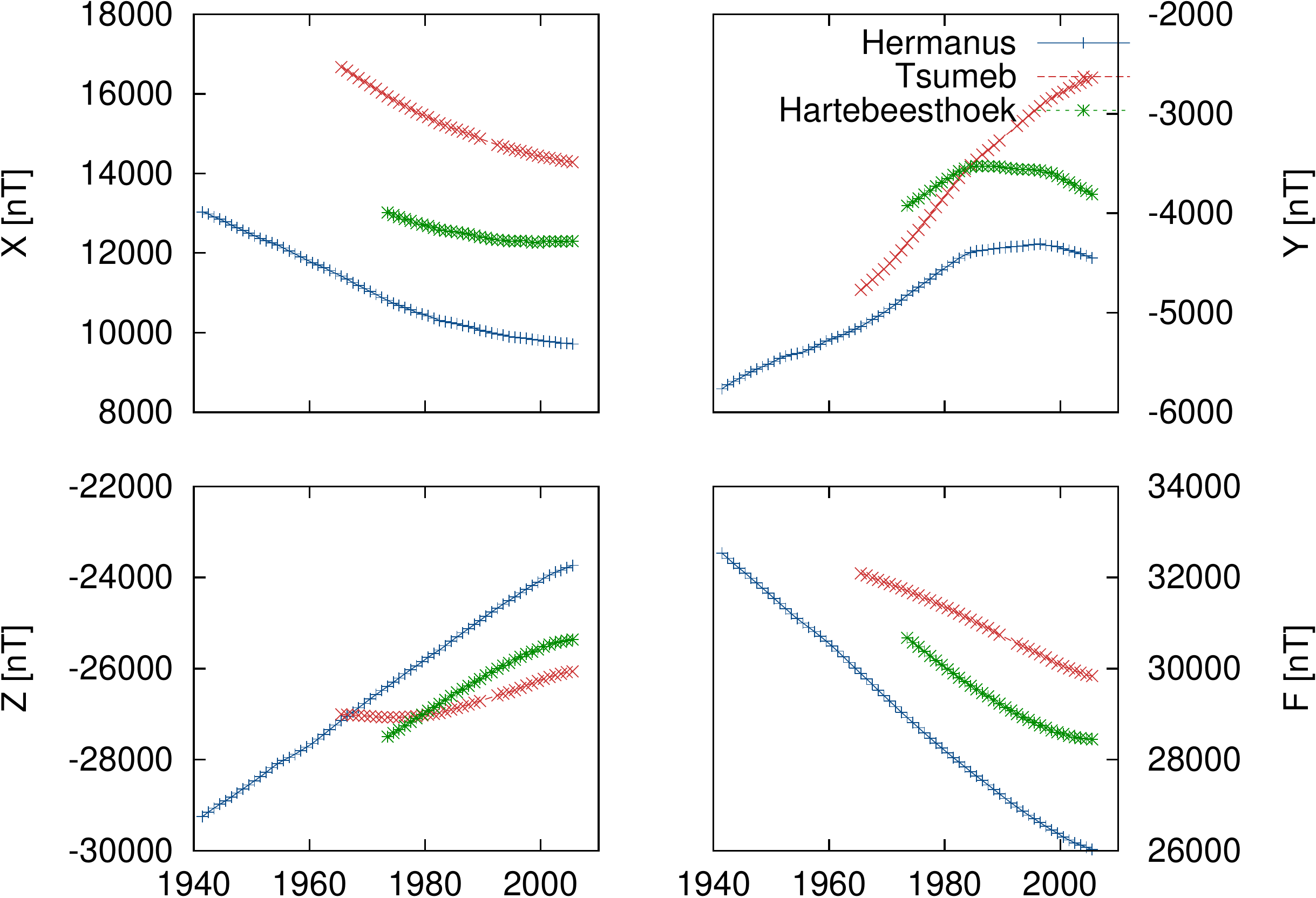}
\caption{Observatory annual mean values for Hermanus (blue), Tsumeb (red) and Hartebeesthoek (green). The figure shows $X$ (North component), $Y$ (East component), $Z$ (vertical down component), as well the total intensity F.}
\label{ann_mean}
\end{figure}
\begin{figure}
\center
\includegraphics[width=8cm,clip]{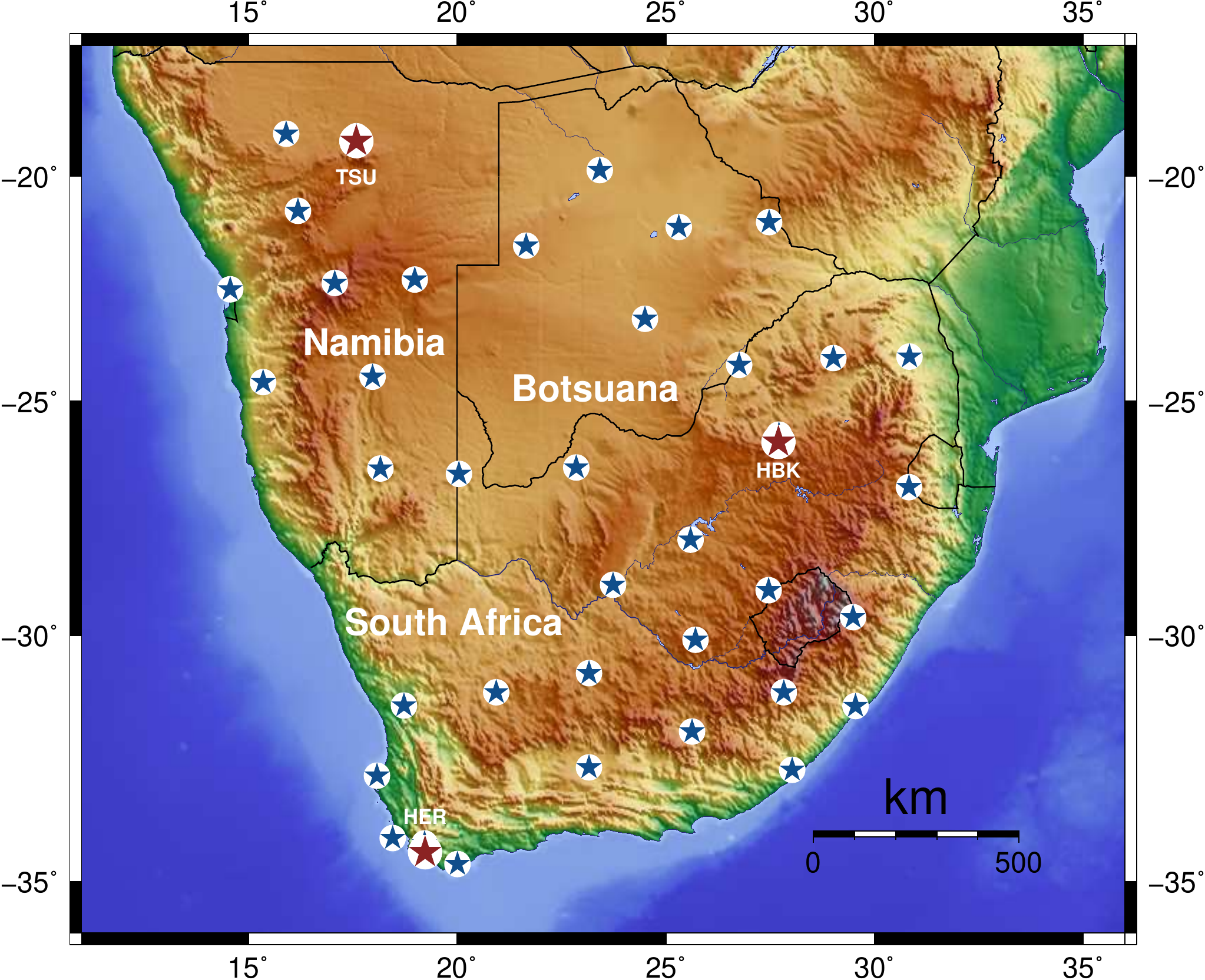}
\caption{Map of distribution of observatories (red) and repeat stations (blue) in the southern African region.  Observatories are situated in Hermanus (HER), Tsumeb (TSU) and Hartebeesthoek (HBK). Not every repeat station has been reoccupied during each survey.}
\label{stations}
\end{figure}
The model we want to produce here is based on ground-based magnetic measurements, provided by the magnetic observatories situated in the southern African region, and also  by the network of repeat stations. Previous to the modelling process, it is crucial to know what kind of preprocessing has been applied to the raw data, and also to investigate the data errors. In the following we present these two classes of data and a comparison with the CM4 model \citep{CM4}. Due to the CM4 validity period, we consider only data available in the time frame between 1961.0 and 2001.0.
\subsection{Data sources}
\subsubsection{Observatory data}
Geomagnetic observatories provide data with an excellent quality, i.e. a low noise level combined with continuous time series. Our region of interest includes three of them that operated between 1961 and 2001: Hermanus (HER), Tsumeb (TSU), and Hartebeesthoek (HBK). Hermanus, close to the southernmost tip of Africa, was established in 1941, Tsumeb in the north of Namibia in 1964, and Hartebeesthoek, north of Johannesburg, in 1971. All these three observatories belong to the INTERMAGNET network and provide high quality  data \cite[]{intermagnet}. The INTERMAGNET demands are met by continuous variation recordings performed by a FGE-type\footnote[1]{Fluxgate magnetometer} magnetometer and baseline determination by means of a DI-flux\footnote[2]{Declination-Inclination Fluxgate magnetometer}. Of course, the data quality decreases back in time, when analog systems were used for variation recordings, and declinometer and QHM \footnote[3]{Quartz Horizontal Magnetometer} for the absolute measurements.
In this study, we include annual mean values from all these three observatories. Altogether, 102 annual mean values are available consisting of 40 vector data  from HER (1961.5-2000.5), 34 vector data  from TSU (1965.5-2000.5, with a gap in 1990/91) and 28 vector data from HBK (1973.5-2000.5). Continuous temporal variations of the field components are plotted in figure \ref{ann_mean}.
\subsubsection{Repeat station data}
Repeat station measurements are conducted to monitor small scale changes in the magnetic field. The stations of the southern African  network are distributed as well as possible all over southern Africa, generally separated by a distance of some $200\mbox{ km}$ from each other. Each single station consists of either a chest-high or a benchmark-like concrete pillar ensuring exact reoccupation. Furthermore, the stations offer mostly an auxiliary narrow pillar for the intensity measurements and, in any case, several azimuth marks.\\
Between 1961 and 2001, the stations were reoccupied on a 5-year basis. Altogether, the network contains more than 100 different stations, of which several had to be closed because of anthropogenetic noise or even vandalism, sometimes because of political reasons or because of lack of funding. We use data only from those stations that were visited at least eight times. To enhance the spatial coverage in the north-east, stations in Botswana are also included as long as they were reoccupied at least five times. These 35 stations used in our study are listed together with the three observatories in table \ref{table_stations}, ordered according to their geomagnetic latitude.\\
\begin{table*}
\center
\begin{tabular}{|r|r|l|r|r|r|r|}
\hline
 station& station &station & Geocentric  & Coordinate  & Geomagnetic  & Coordinate\\
 number & code    &name    & colatitude  & longitude   & colatitude   & longitude \\
\hline
\hline
 1 &  OKA &       Okaukuejo & 109.03 &  15.91 & 108.25 & 83.95 \\ 
 2 &  TSU &          Tsumeb & 109.08 &  17.58 & 108.60 & 85.58 \\ 
 3 &  KAL &        Kalkfeld & 110.77 &  16.18 & 110.01 & 83.86 \\ 
 4 &  MAU &            Maun & 109.85 &  23.42 & 110.41 & 91.15 \\ 
 5 &  SWA &      Swakopmund & 112.53 &  14.57 & 111.44 & 81.93 \\ 
 6 &  WIN &        Windhoek & 112.39 &  17.07 & 111.76 & 84.40 \\ 
 7 &  GHA &          Ghanzi & 111.56 &  21.66 & 111.77 & 89.06 \\ 
 8 &  ORA &           Orapa & 111.14 &  25.31 & 112.02 & 92.75 \\ 
 9 &  GOB &         Gobabis & 112.32 &  18.99 & 112.03 & 86.28 \\ 
10 &  FRA &     Francistown & 111.03 &  27.48 & 112.30 & 94.92 \\ 
11 &  SOS &      Sossusvley & 114.59 &  15.35 & 113.60 & 82.25 \\ 
12 &  KHU &          Khutse & 113.19 &  24.50 & 113.88 & 91.51 \\ 
13 &  MAR &       Mariental & 114.46 &  17.97 & 113.94 & 84.83 \\ 
14 &  MCO &          Marico & 114.20 &  26.75 & 115.29 & 93.51 \\ 
15 &  POT &   Potgietersrus & 114.05 &  29.02 & 115.55 & 95.79 \\ 
16 &  MIC &            Mica & 114.02 &  30.84 & 115.84 & 97.61 \\ 
17 &  KMH &    Keetmanshoop & 116.46 &  18.16 & 115.94 & 84.58 \\ 
18 &  RIF &     Rietfontein & 116.58 &  20.04 & 116.40 & 86.39 \\ 
19 &  SEV &          Severn & 116.44 &  22.86 & 116.77 & 89.18 \\ 
20 &  HBK &  Hartebeesthoek & 115.73 &  27.71 & 116.96 & 94.12 \\ 
21 &  PIT &     Piet Retief & 116.86 &  30.82 & 118.63 & 96.96 \\ 
22 &  HTZ &    Hertzogville & 117.99 &  25.59 & 118.78 & 91.51 \\ 
23 &  DOU &         Douglas & 118.94 &  23.74 & 119.37 & 89.46 \\ 
24 &  LAD &       Ladybrand & 119.05 &  27.46 & 120.16 & 93.11 \\ 
25 &  SPR &   Springfontein & 120.09 &  25.71 & 120.86 & 91.13 \\ 
26 &  VAN &    Vanrhynsdorp & 121.45 &  18.73 & 120.93 & 83.96 \\ 
27 &  WIL &       Williston & 121.18 &  20.94 & 121.06 & 86.18 \\ 
28 &  FON &       Fontentje & 120.78 &  23.15 & 121.07 & 88.44 \\ 
29 &  UND &       Underberg & 119.62 &  29.49 & 121.08 & 94.98 \\ 
30 &  LAN &       Langebaan & 122.89 &  18.08 & 122.21 & 82.97 \\ 
31 &  ELL &          Elliot & 121.18 &  27.83 & 122.31 & 92.95 \\ 
32 &  CRA &         Cradock & 121.99 &  25.63 & 122.70 & 90.58 \\ 
33 &  POR &  Port St. Johns & 121.46 &  29.54 & 122.89 & 94.58 \\ 
34 &  RIB &        Rietbron & 122.72 &  23.15 & 122.96 & 87.96 \\ 
35 &  BUF &     Buffels Bay & 124.13 &  18.45 & 123.50 & 83.01 \\ 
36 &  HER &        Hermanus & 124.25 &  19.23 & 123.74 & 83.73 \\ 
37 &  GON &   Gonubie Mouth & 122.76 &  28.03 & 123.89 & 92.74 \\ 
38 &  AGU &         Agulhas & 124.65 &  20.00 & 124.28 & 84.38 \\ 
\hline
\end{tabular}
\label{table_stations}
\caption{Geomagnetic observatories and repeat stations considered in the present study.}
\end{table*}
\normalsize
The data resulting from a repeat station survey give comparable information to observatory annual mean values, but with a reduced accuracy. The sources of errors are manifold. Firstly, the conditions in the field are never as good as in an observatory: the weather has its impact on the performance of the measurement procedure, tripods are less stable than concrete pillars and if instruments fail, they generally cannot be repaired on-site. Secondly, the data measured at a given time in the year are always reduced to an epoch, or at the middle of the year, using the closest magnetic observatory. The estimated field component $E(t_e)$ at the repeat station is 
\footnotesize
\[
E_{FS}(t_e)=E_{Obs}(t_e)-E_{Obs}(t_s)+E_{FS}(t_s)
\]
\normalsize
where $t_s$ denotes the epoch of the survey, $t_e$ the epoch to be reduced to and $E_{Obs}$, $E_{FS}$ are the field components at the control observatory and the field station, respectively. A detailed description of the handling of repeat station data can be found in \cite{NewittEtAl1996}.
\begin{figure*}
\center
\includegraphics[width=7cm,angle=270]{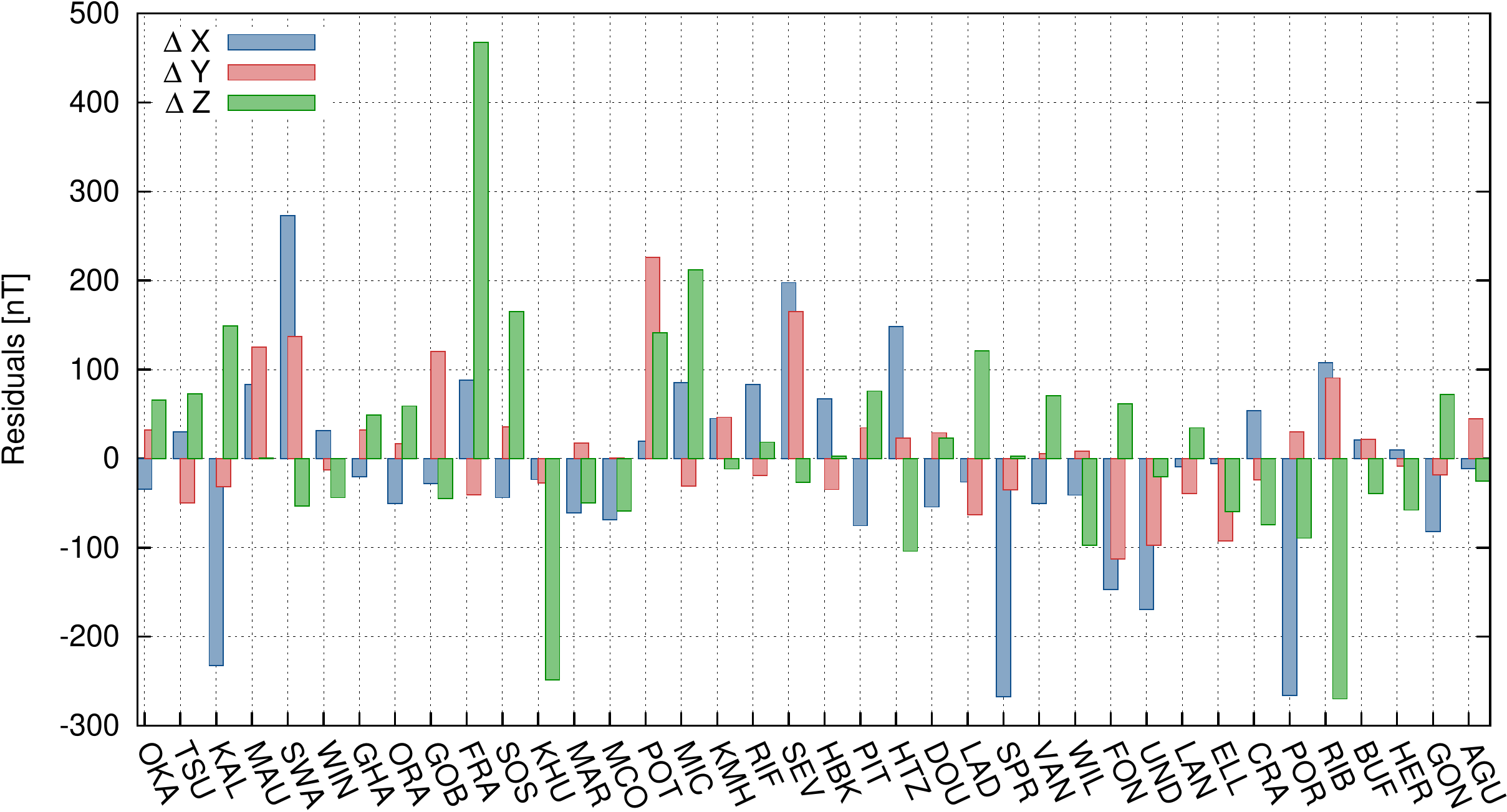}
\caption{Crustal biases for observatory and repeat station locations calculated as differences between annual mean values based on CM4  model and annual mean values of given measurement points.}
\label{biases}
\end{figure*}
\subsection{Real and synthetic data comparison}
\label{sec:2.2}
\begin{figure}
\center
\includegraphics[height=21cm]{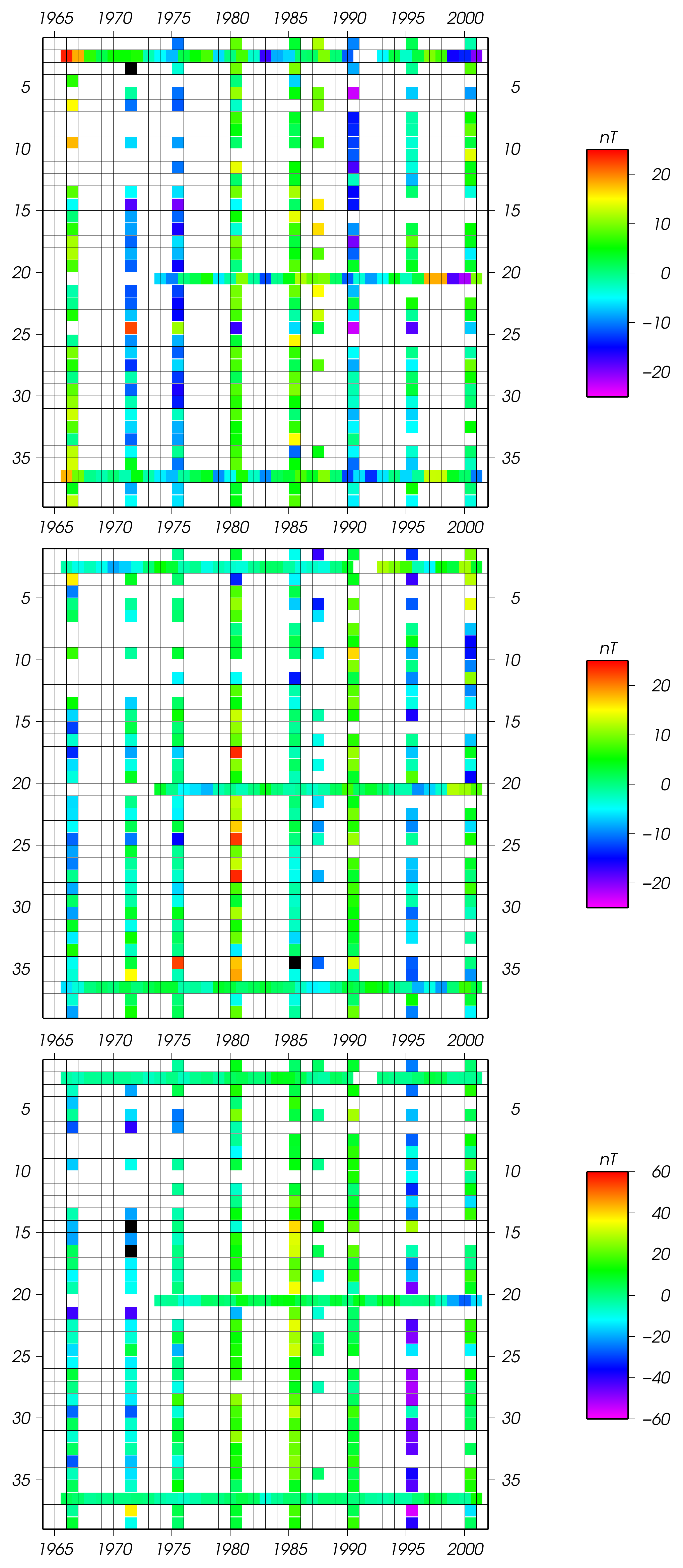}
\caption{Differences between the real data and the corresponding synthetic data estimated from CM4 model, ordered according to the station and observatory numbers (cf. table \ref{table_stations}). Crustal biases (average of differences between data and CM4 predictions for each location) are subtracted.  $X$, $Y$ and $Z$ components are shown from top to bottom. Variations in time are similar for all stations. The graph on the bottom displays the sunspot number (blue) and the solar flux irradiation (red) as an expression of external field activity.}
\label{picto}
\end{figure}

In order to model the magnetic measurements, some preprocessing steps are crucial.
Firstly, every single time series is checked in order to detect obvious jumps. This task can be done by comparing the available annual means with a synthetic data set, computed for the same locations and the same epochs from a global magnetic field model. Considering the time period covered by the available data, we have selected the CM4 model \citep{CM4}, including core and lithospheric field up to maximum degree 65. For each  data point in space and time, where a measurement is available, we calculated the corresponding synthetic annual mean value and compared the result to the real data. The general behaviour is the same for real and synthetic data, but for all stations and all components, a bias remains ranging from a few up to some hundreds of nT. These biases between measured and synthetic data are averaged over the whole time span and regarded as unmodelled crustal contribution. For all stations used  in this study, these crustal biases are shown in figure \ref{biases}. We have compared our results to the observatory biases described in \cite{MandeaLanglais2002} and to the GAMMA candidate model for the World Digital Magnetic Anomaly Map \citep{WDMAM}, without finding however a good agreement. \\
We subtract CM4 annual mean estimates and crustal biases (as given in fig. \ref{biases}) from the data. The remaining quantities do not stay constant, but vary slightly in time. Ordered and numbered according to geomagnetic latitude (cf. table \ref{table_stations}), these differences are plotted in figure \ref{picto}, for $X$, $Y$ and $Z$ components. The sun spot number and the solar flux are also shown. Due to the fact that the repeat stations were generally visited only  every five years, the matrix of the temporal variation is scarcely filled; only observatory data represent continuous time series. Still, some common features can be noticed:  for $Y$ and $Z$, the differences are negative in the solar minimum in 1995 and positive in the solar maximum in 1990, and vice versa for $X$. This trend can be discovered also for the other years, but less distinctively.  Overall we think this is a weak indication that external field influences are contributing to the differences.\\
For European observatory annual means, \cite{VerbanacEtAl2007} proposed a correction scheme for external contributions. In the present study, this approach cannot be applied, as the obtained matrix of the temporal evolution is too sparse.\\
The standard deviation for each component is regarded as an estimate of the noise in the data. While for the observatory data the deviations amount only to $\pm 5 \mbox{ nT}$, the deviations at the  field stations reach up to $\pm 15 \mbox{ nT}$ in all components. For all time series, this standard deviation is used later on to estimate elements of the covariance matrix.
\section{Harmonic splines}\label{method}
Our aim is to build a regional model of the magnetic field over the southern African  region using harmonic splines. Our method goes back to \cite{ShureEtAL1982} where a detailed description can be found. Highlighting localised constraints connected with local base functions, \citet{Lesur2006} describes a representation close to the method used here. We first develop the spatial description of the field and introduce a time dependency later on.
\subsection{Theory}
On the (source-free) surface of the Earth, the magnetic field $\textbf{B}$ can be expressed as the gradient of a 
scalar potential $\Phi$, with $\textbf{B} = -\nabla \Phi$ and $\nabla^2\Phi=0$. Using the 
Spherical Harmonics (SH) $Y_l^m(\vartheta,\varphi)$ of degree $l$ and order $m$, the radial component of the internal 
part of the global magnetic field can be written as:
\begin{equation}
B_r=\sum^{L}_{l=0}\sum_{m=-l}^lg^{m}_{l}(l+1) \left(\frac{a}{r}\right)^{l+2} Y^{m}_{l}(\vartheta,\varphi)\; ,
\end{equation}
where $\vartheta,\varphi$ are the geocentric colatitude and longitude, the $g^{m}_{l}$ are the Gauss coefficients, 
and $a$ is a reference radius ($6371.2\mbox{ km}$). $Y_l^m(\vartheta,\varphi)$ are the usual Schmidt semi-normalised SH functions. We use here the convention that negative orders
($m < 0$) are associated with $\sin (m\varphi)$ terms, whereas zero or positive orders ($m \geq 0$) are associated 
with $\cos (m\varphi)$.\\
We consider a functional $\phi$ as the integral of the squared second horizontal divergence of  $B_r$ over a sphere $\Omega$ 
of radius $a$:
\begin{equation} \label{functional}
\phi=\int_\Omega |\nabla^2_h B_r|^2\mathrm{d}\omega .
\end{equation}
Exploiting the orthogonality of the spherical harmonics we obtain:
\begin{equation}
\phi=\sum^{L}_{l,m}4\pi\frac{ l^2(l+1)^4}{2l+1} (g^m_l)^2 ;
\end{equation}
where  $\sum^{L}_{l,m}$ denotes the double sum $\sum^{L}_{l=1}\sum^{l}_{m=-l}$ .
Accordingly, we define a scalar product:
\begin{equation}
<\mathbf{a},\mathbf{b}>=\sum^{L}_{l,m}4\pi\frac{ l^2(l+1)^4}{2l+1}a^m_l b^m_l.
\label{eq:3.1}
\end{equation}
We now consider a data set made of $N$ measurements of the radial component of the magnetic field at given points 
$(\vartheta_i, \varphi_i, r_i)$, $i=1,2, \ldots ,N$. We write each of these values as a scalar product of the form (\ref{eq:3.1}).
\begin{eqnarray} \label{B_r}
\nonumber {B_r(\vartheta_i, \varphi_i, r_i)}_{}&=&\sum^{L}_{l,m}(l+1) \left(\frac{a}{r_i}\right)^{l+2} Y^m_l(\vartheta_i, \varphi_i)g^m_l\\ 
\nonumber         &=&\sum^{L}_{l,m}4\pi\frac{(l+1)^4l^2}{2l+1}\;k^m_{li}\; g^m_l\\ 
	 &=&<\mathbf{k_i},\mathbf{g}>
\label{eq:3.2}
\end{eqnarray}
where $\mathbf{k_i}=[\;k^m_{li}\;]_{\{l,m\}}$ and $\mathbf{g}=[\;g^m_{l}\;]_{\{l,m\}}$. The $ k^m_{li} $ are defined by:
\begin{equation}
k^m_{li}= \frac{2l+1}{4\pi(l+1)^3l^2}\left(\frac{a}{r_i}\right)^{l+2} Y^m_l(\vartheta_i, \varphi_i).
\end{equation}
The $\mathbf{k_i}$ are linearly independent vectors \cite[]{Parker1977}, and therefore, $\mathbf{g}$ can be written as a linear combination of these vectors:
\begin{equation}
\mathbf{g}=\sum_{j=1}^{N} \alpha_j^r \mathbf{k_j} .
\label{eq:3.3}
\end{equation}
Writting equation (\ref{eq:3.2}) for all $N$ data values and using equation (\ref{eq:3.3}) leads to the linear system:
\begin{equation}
[B_r(\vartheta_i, \varphi_i, r_i)]_{\{i\}}= \mathbf{\Gamma} \cdot \mathbf{\alpha} \mathbf{^r}
\end{equation}
where $\mathbf{\alpha^r} = [\;\alpha_j^r\;]_{\{ j \}}$ and the Gram matrix $\mathbf{\Gamma}$ is defined by:
\begin{equation}
\Gamma_{ij}=\sum^L_{l,m}4\pi\frac{(l+1)^4l^2}{2l+1}k^m_{li}k^m_{lj} .
\end{equation}
Once the $\mathbf{\alpha^r}$ vector has been estimated, the magnetic field can be computed at any point $(\vartheta, \varphi, r)$ by: 
\begin{eqnarray} 
\nonumber\mathbf{B}(\vartheta, \varphi, r) &=& -\nabla a \sum^L_{l,m}g^m_l\left(\frac{a}{r}\right)^{l+1}Y^m_l(\vartheta, \varphi)\\ 
\nonumber&=& -\nabla a \sum^L_{l,m} \left\{\sum_j k^m_{lj}\alpha_j^r\right\}\left(\frac{a}{r}\right)^{l+1}Y^m_l(\vartheta, \varphi)\\ 
\nonumber&=& -\nabla \sum_j \alpha_j^r \left\{a\sum^L_{l,m} \frac{2l+1}{4\pi l^2(l+1)^3} \right. \times \\ &&\left.\left(\frac{a}{r_j}\right)^{l+2}Y^m_l(\vartheta_j, \varphi_j)\left(\frac{a}{r}\right)^{l+1} Y^m_l(\vartheta, \varphi)\right\}
\end {eqnarray}
We rewrite this expression in a compact form:
\begin{equation}
\mathbf{B}(\vartheta, \varphi, r)=-\sum_j \alpha_j^r \;\nabla F^{Lr}_j(\vartheta, \varphi, r) ,
\label{eq:3.4}
\end{equation}
where the function $F^{Lr}_j(\vartheta, \varphi, r)$ is defined by:
\begin{equation} 
F^{Lr}_j=a\sum^L_{l,m}f_l(l+1)
\left[\left(\frac{a}{r_j}\right)^{l+2} Y^m_l(\vartheta_j, \varphi_j)\right]
\left(\frac{a}{r  }\right)^{l+1} Y^m_l(\vartheta, \varphi) ,
\label{eq:3.5}
\end{equation}
with $f_l=\tfrac{2l+1}{4\pi (l+1)^4l^2}$. The benefit and special characteristic of the derived coefficient $f_l$ 
is that the obtained field model minimises the functional $\phi$ declared in equation (\ref{functional}). The functions $F^{Lr}_j$ are defined at the measurement positions ($\vartheta_j, \varphi_j,r_j, j=1,...,N$) and are interpolatory \cite[]{Parker1977}, i.e. they allow an exact fit to the radial component data, as long as the data contain SH degree lower than or equal to $L$.\\
So far we have been treating radial component data, only. However the magnetic field model defined in equation 
(\ref{eq:3.4}) might not fit the tangential components of the magnetic field at the points $(\vartheta_i, \varphi_i, r_i)$, $i=1,2, \ldots ,N$. To include them in the model derivation, we have to introduce two further series of functions, similar to $F^{Lr}_j$ in equation (\ref{eq:3.5}):
\begin{eqnarray} \nonumber
F^{L\vartheta}_j(\vartheta, \varphi, r)&=&a\sum^L_{l,m} f_l
\left(\frac{a}{r_j}\right)^{l+2}\partial_\vartheta Y^m_l(\vartheta_j, \varphi_j)\\&&\times
\left(\frac{a}{r}\right)^{l+1} Y^m_l(\vartheta, \varphi),
\label{eq:3.6}
 \end{eqnarray}
\begin{eqnarray}\nonumber
F^{L\varphi}_j(\vartheta, \varphi, r)&=&a\sum^L_{l,m} f_l
\left(\frac{a}{r_j}\right)^{l+2}\frac{\partial_\varphi}{\sin \vartheta_j}Y^m_l(\vartheta_j, \varphi_j)\\&&\times
\left(\frac{a}{r}\right)^{l+1} Y^m_l(\vartheta, \varphi).
\label{eq:3.7}
\end{eqnarray}
The derivations of $F^{L\vartheta}_j$ and $F^{L\varphi}_j$ are given in  appendix \ref{a:add_kernels}.\\
The magnetic field can then be written as:
\begin{equation}\label{eq:3.20}
\mathbf{B}(\vartheta,\varphi,r)=- \nabla\left(
   \sum_j\alpha_j^r F^{Lr}_j(\vartheta, \varphi, r)
  +\sum_j\alpha_j^\vartheta F^{L\vartheta}_j(\vartheta, \varphi, r)
  +\sum_j\alpha_j^\varphi  F^{L\varphi  }_j(\vartheta, \varphi, r)\right)
\end{equation}With this representation, and because the functions are defined at the observation locations, we are now able to fit three component data exactly, as long as the data contain SH degrees lower than or equal to $L$.\\
This method  is equivalent to the harmonic spline method presented in \cite{ShureEtAL1982}. The main difference comes from the use of finite series to define $F^{L r}_j,F^{L\vartheta}_j$ and $ F^{L\varphi}_j$. We point out that the choice of a truncated series allows them to be included in a global field model, itself made of spherical harmonics.\\
If eq. (\ref{eq:3.20}) is time independent, it defines a linear system of equations where the $\alpha_j^r$, $\alpha_j^\vartheta$ and  $\alpha_j^\varphi$ are the unknowns, the data are magnetic field measurements and the matrix is made of combinations of the functions $^{Lr}_j$, $F^{L\vartheta}_j$ and $F^{L\varphi  }_j$. This linear system is square and can be solved directly. However, here we introduce a time dependency and we expand each of the coefficients $\alpha_j^r$ on a basis of B-Splines, i.e. piecewise polynomials between spline knots:
\begin{equation}
 \alpha_j^{r}=\sum_{k=1}^{n_{knots}}\beta_{kj}^{r}b_k(t)
\label{eq:3.8}
\end{equation}
and similarly for the $\alpha_j^\varphi$ and $\alpha_j^\vartheta$.\\
\subsection{Implementation}
For time representation, we use order six B-splines with 27 spline knots spaced at 2.5 year intervals from 1961.0 to 2001.0. The maximum SH degree $L$ in equations (\ref{eq:3.5}), (\ref{eq:3.6}) and (\ref{eq:3.7})  is set to 20 as a result of our study on synthetic data (see next section). This degree is small enough to show mostly contributions from the core, but still large enough to resolve the smaller scale features.\\
We solve the linear system with the classical least squares. A linear system $\mathbf{d}=\mathbf{Am}$ is built from equations (\ref{eq:3.20}) and (\ref{eq:3.8}). We solve for the unknown model parameters $\beta_{kj}^{r}$, $\beta_{kj}^{\vartheta}$ and $\beta_{kj}^{\varphi}$ as defined in eq. (\ref{eq:3.8}) and for the crustal offsets at each measurement location. The $\beta_{kj}^{r}$, $\beta_{kj}^{\vartheta}$ and $\beta_{kj}^{\varphi}$ together with the crustal offsets constitute the parameter vector $\mathbf m$. In a least squares sense, the solving is done by minimising the functional $J$:
\begin{equation}
\label{minim}
 J=(\mathbf{d}-\mathbf{Am})^T\textbf C_e^{-1}(\mathbf{d}-\mathbf{Am})
\end{equation}
 The elements of the covariance matrix $\mathbf C_e^{-1}$ are obtained for each station and each component from the standard deviation of the scatter of the data about the mean difference to the CM4 model prediction (see section \ref{sec:2.2}).\\To obtain a robust model, regularization is required. For the regularization in time we minimise the squared second time derivative of the radial field integrated over the whole sphere and the whole timespan:
\begin{eqnarray}\label{int}
&&\int_{t=t_{start}}^{t=t_{end}}\int_\Omega (\partial^2_t B_r)^2 d\omega dt
\end{eqnarray}
which can be written as 
\begin{eqnarray}
 \int_{t=t_{start}}^{t=t_{end}} \int_\Omega \left[(\ddot B_r^r)^2+(\ddot B_r^\vartheta)^2+(\ddot B_r^\varphi)^2\right.
+ \;2\ddot B_r^r\ddot B_r^\vartheta  \left. + 2\ddot B_r^r\ddot B_r^\varphi  + 2\ddot B_r^\vartheta\ddot B_r^\varphi \right ]  d\omega dt .
\end{eqnarray}
The expressions for the six integrals are given in appendix \ref{a:int}.\\
For the regularization in space we minimise the integral over space and time of the squared horizontal divergence of the  radial field component:
\begin{equation}
  \int_{t=t_{start}}^{t=t_{end}}\int_\Omega | \nabla^2_h B_r|^2 d\omega dt .
\end{equation}
Accordingly, damping matrices  $\Lambda_s$ and $\Lambda_t$ are introduced  for regularization in space and time, respectively.
Thus, we rewrite equation (\ref{minim}) as
\begin{equation}
 J=(\mathbf{d}-\mathbf{Am})^T\textbf C_e^{-1}(\mathbf{d}-\mathbf{Am})+\lambda_t \textbf m ^T \Lambda_t \textbf m
+\lambda_s \textbf m ^T \Lambda_s \textbf m
\end{equation}
The choice of $\lambda_t$ and $\lambda_s$ is discussed in the next section.
\section{Results}
In this section, we describe the tests performed on synthetic data, giving us the possibility to estimate important parameters ($\lambda_s$, $\lambda_t$, $L$). Thereafter, we compute a main field model and investigate its changes in time.
\subsection{Tests with synthetic data}
\begin{figure}
 \includegraphics[width=7cm]{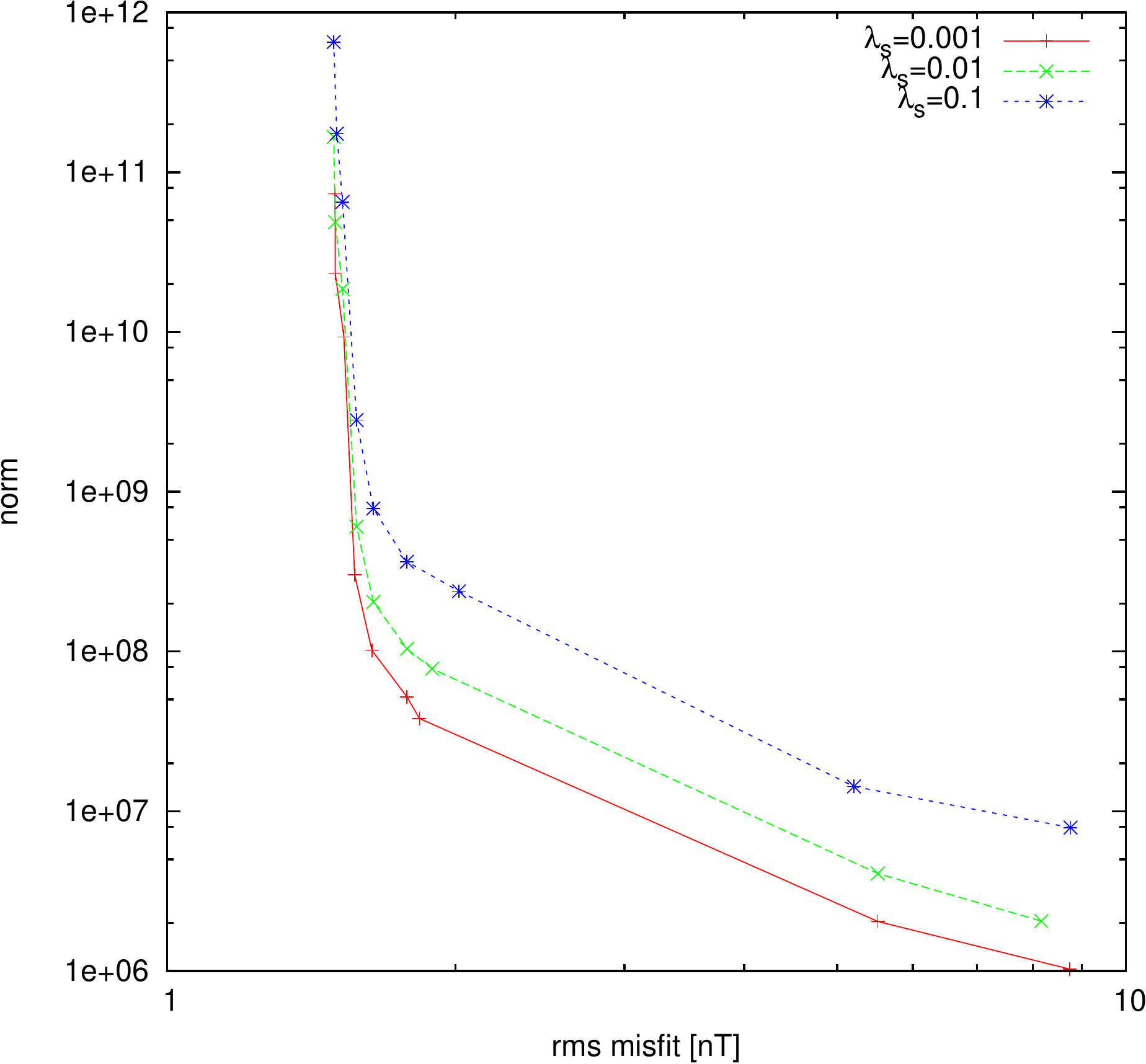}
 \caption{Trade-off curves between data misfit and the roughness of the model. The three lines show trade-off curves for three fixed $\lambda_s$ ($\lambda_s=0.001$ in red, $\lambda_s=0.01$ in green and $\lambda_s=0.1$ in blue) and varying $\lambda_t$ ($\lambda_t=1e^{-1},...,1e^{-9}$.) The norm was calculated according to eq. (\ref{int})}.
  \label{tradeoff}
 \end{figure}
The spline modelling approach has been  thoroughly tested before being applied to real data. During the testing phase some important parameters in the modelling have been estimated:
\begin{itemize}
\item the maximum degree $L$,
\item the damping parameters $\lambda_s$ and $\lambda_t$ in space and time.
\end{itemize}
We use the noise-free synthetic data set derived from CM4 as input. As described in section \ref{data}, a synthetic annual mean value was computed for each  data point in space and time where a real measurement is available. Using this synthetic data set, we first change $L$ from 5 to 60 without imposing any damping and see that we can fit the synthetic data satisfactorily, i.e. to a rms in the order of $0.1 \mbox{ nT}$, using a maximum degree $L=20$.\\
As a next step, we vary the damping parameters $\lambda_s$ and $\lambda_t$ and calculate the associated norms. Figure \ref{tradeoff} shows several trade-off curves for three fixed $\lambda_s$ and varying $\lambda_t$. A knee is prominent for $\lambda_t=1 \times 10^{-4}$ in all three curves. The same has been done for fixed $\lambda_t$ and varying $\lambda_s$ (not shown here) leading to  $\lambda_s=1 \times 10^{-1}$.\\
\subsection{Regional Modelling}
Applying the spline technique with the previously indicated values for $\lambda_s$ and $\lambda_t$ to the real data set allows us to obtain a temporally varying core field model for the region of interest. From this, we also derive the secular variation.
\subsubsection{Core Field Modelling}
\begin{figure*}
\center
 \includegraphics[width=16cm]{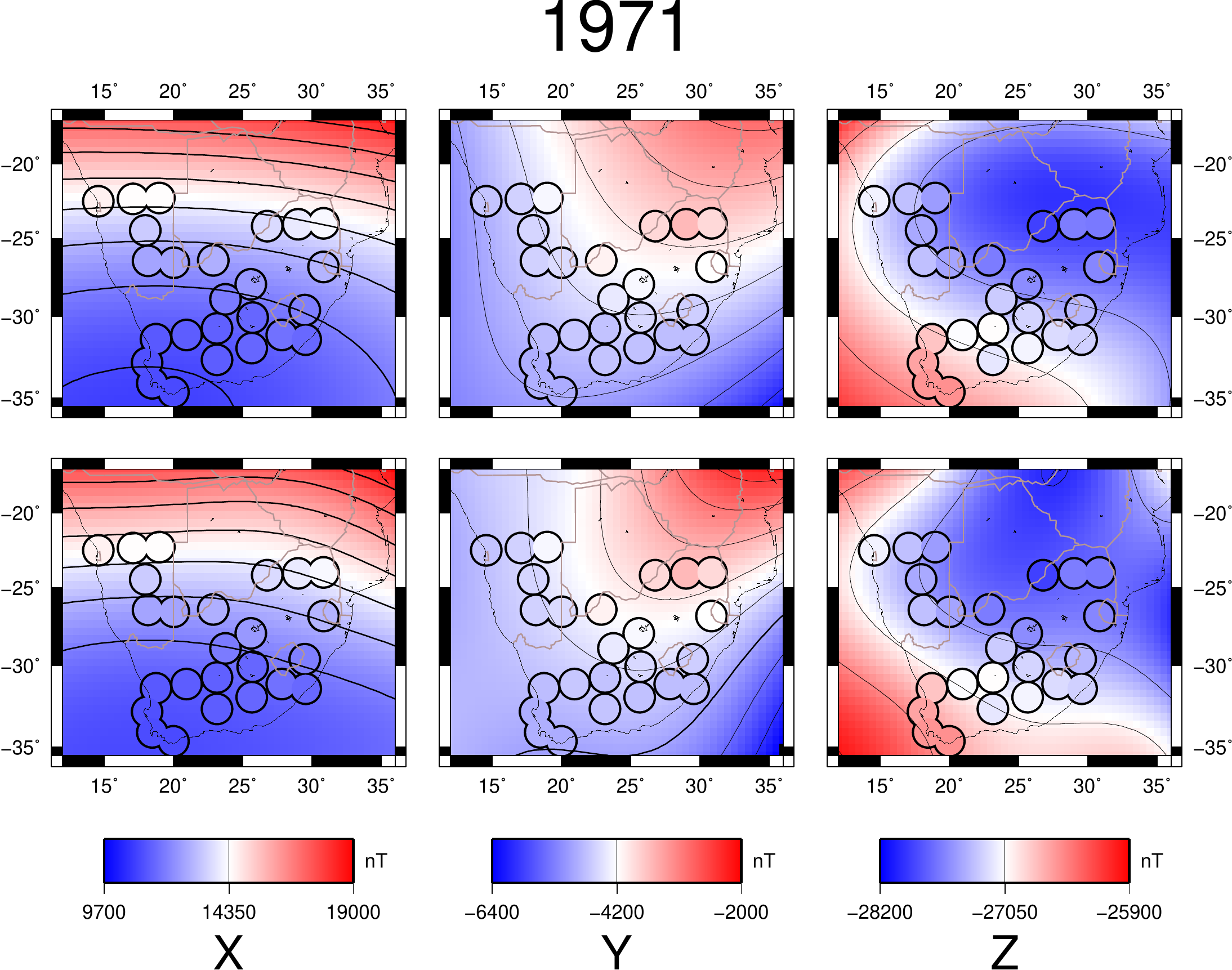}
 \includegraphics[width=16cm]{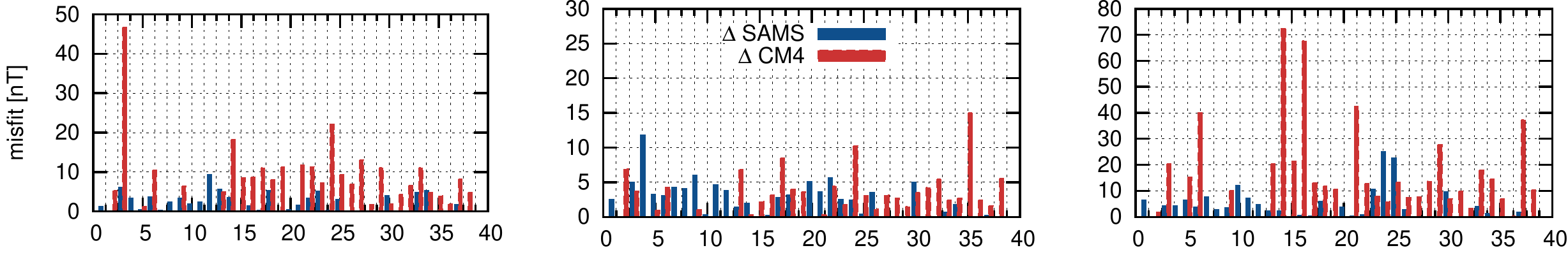}
 \caption{Maps of the $X$, $Y$ and $Z$ components of the core field in (nT) as derived from CM4 (top) and  SAMS (bottom) models for the epoch 1971. Each disc, centred on an observatory or repeat station position, indicates the measured data at this epoch. The same colorscales are used for models and measurements. Contour lines are separated by $1000 \mbox{ nT}$ for X and by $500 \mbox{ nT}$ for Y and Z. The bottom panels show the differences (for $X$, $Y$ and $Z$ component, with different vertical scales) between  measurements and synthetic data derived from SAMS (blue). For comparison, the differences between measurements and synthetic data derived from CM4 corrected for the crustal offset as estimated in section \ref{sec:2.2} are also shown. The station and observatory numbers correspond to those in table \ref{table_stations}}
\label{1971}
\end{figure*}

\begin{figure*}
\center
 \includegraphics[width=16cm]{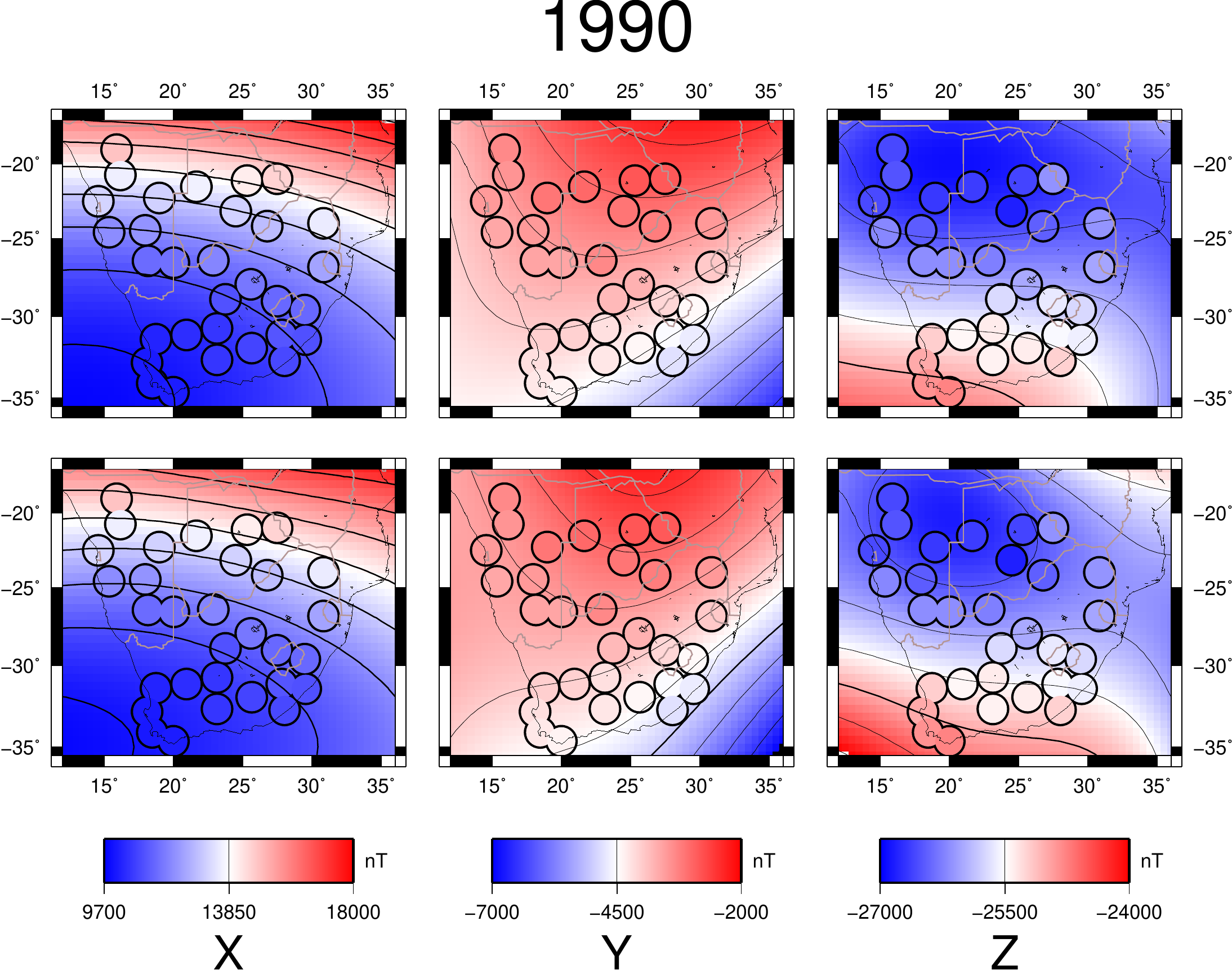}
 \includegraphics[width=16cm]{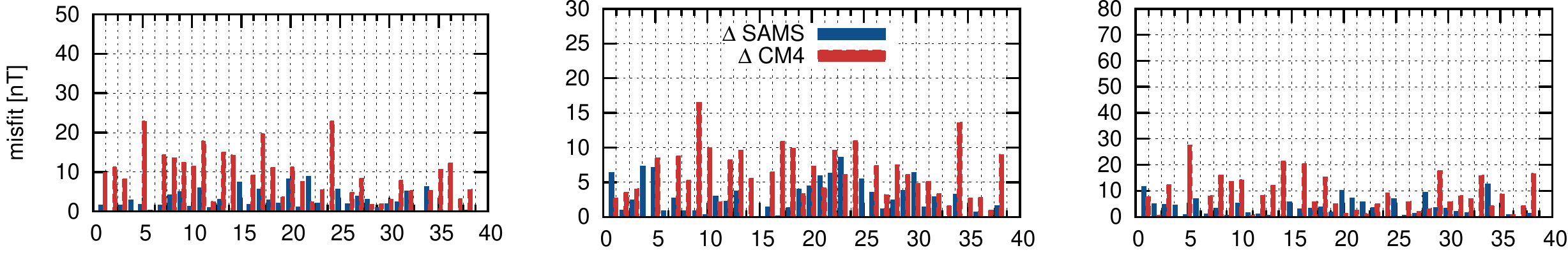}
 \caption{Same as figure \ref{1971} for the epoch 1990.}
\label{1990}
\end{figure*}

The modelling scheme has been applied to the data set described in section \ref{data}. The quality of the obtained Southern African Model made of Splines (SAMS) is firstly checked by comparing the residuals with respect to the data. We compute the residual average for all three components. It stays below $0.01\mbox{ nT}$ with a rms value of about $4.5\mbox{ nT}$ for $X$ and $Y$ components  and $6.5\mbox{ nT}$ for the $Z$ component. This is of the order of the expected data error.\\
From the SAMS model, we can compute synthetic data at any epoch. However, we choose to show here results for 1971 and 1990, epochs for which measurements are available for nearly all stations. Figures \ref{1971} and \ref{1990} show comparative maps for the core field derived from the SAMS and CM4 models. For this, the measurements reported for a given epoch and for a given location are also included using the same colorscale. The regional SAMS model fits the available measurements better than CM4 does. Moreover, SAMS is able to describe some small scale features, such as the curvature  in the $Z$ component, also visible in $X$ and  $Y$. To underline this fact, we show the differences separately, ordered according to the station location. These differences, generally of the order of some tens of nT for the 1990 epoch, are smaller when SAMS is used. For both models, the residuals are dramatically reduced for the Z-component for 1990 compared to 1971, indicating clearly the change in the instrumentation used in field measurements.\\
\begin{figure*}
 \includegraphics[width=7cm]{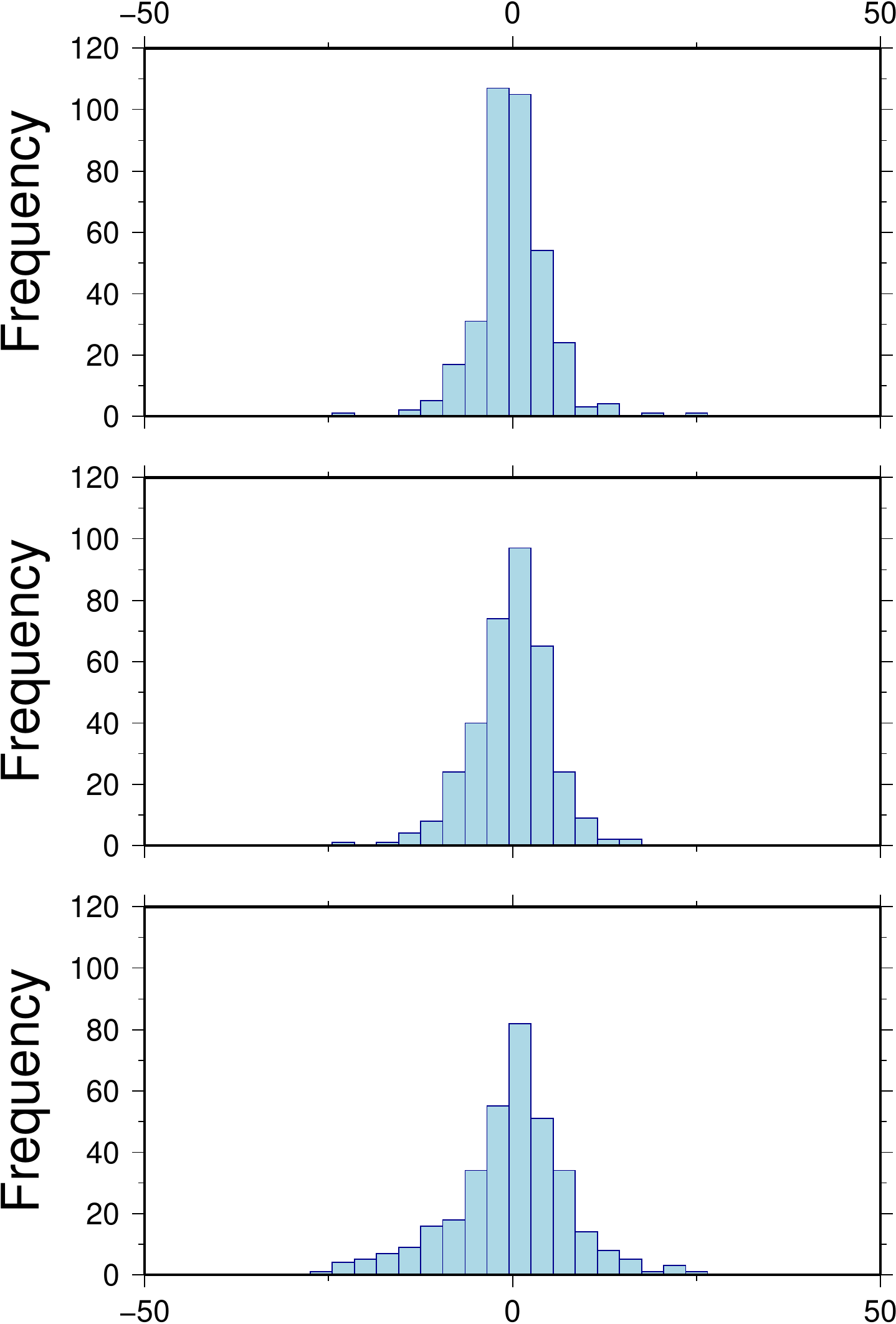}
\hspace{3cm}
 \includegraphics[width=7cm]{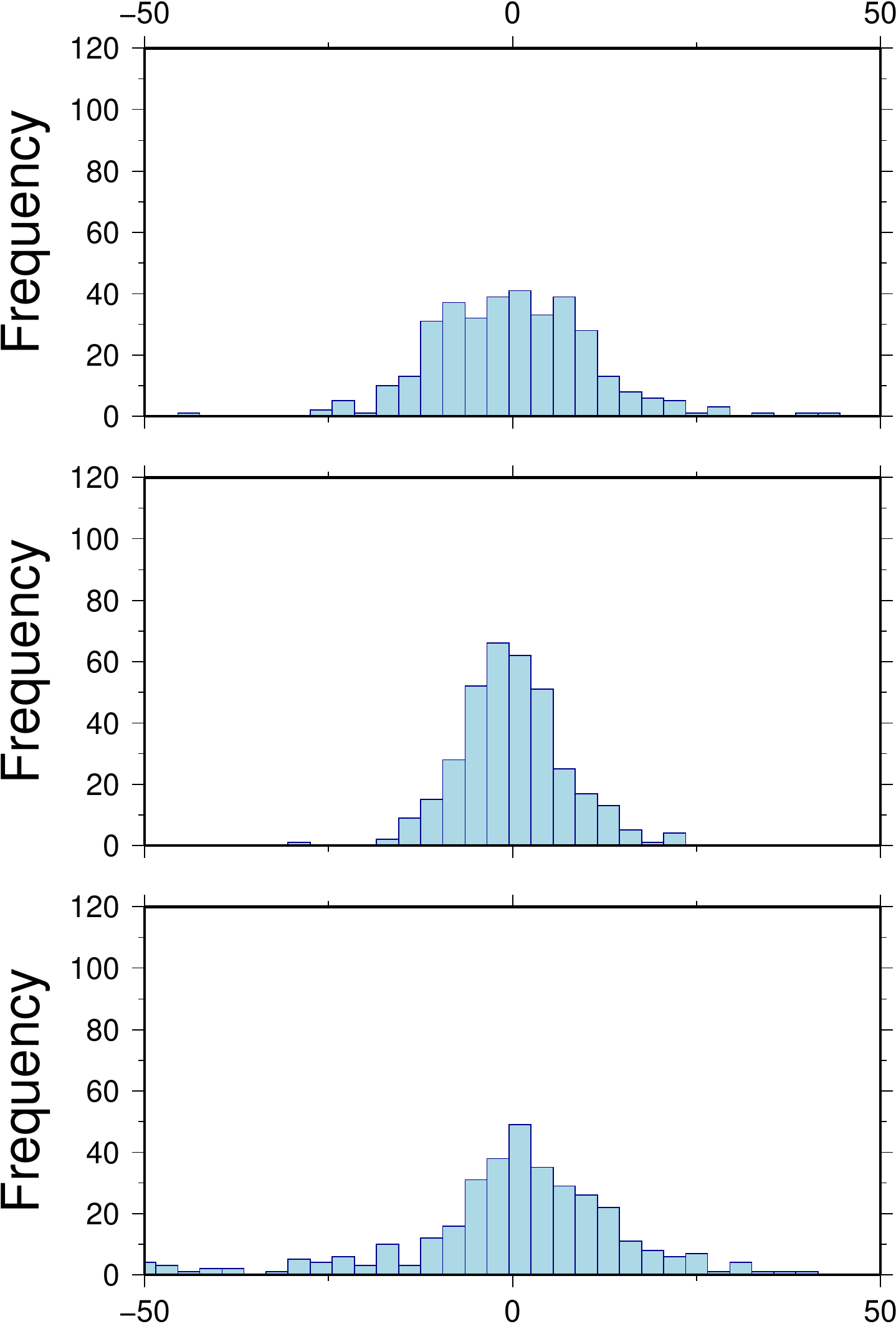}
 \caption{Histograms of misfit for all stations and all years when modeled by SAMS (top) and CM4 (bottom) models have been used.   The $X$, $Y$, and $Z$  components are shown from top to bottom.}
\label{histo2}
\end{figure*}
Finally, we compute histograms of misfit when data are predicted by SAMS and CM4 (figure \ref{histo2}). The histograms obtained from SAMS are narrower, centred on zero and decrease rapidly, while those computed from CM4 are much broader.
\subsubsection{Secular Variation}
As for the core field, we compute and draw maps of the $X$, $Y$, and $Z$ secular variation. They are available as animations\footnote{see supplementary material: http://www.gfz-potsdam.de/portal/gfz/ Struktur/Departments/Department+2/sec23/projects/modeling/SAMS} and can be  used to identify the time and location of maximum variations.
\begin{figure*}
 \includegraphics[width=7cm]{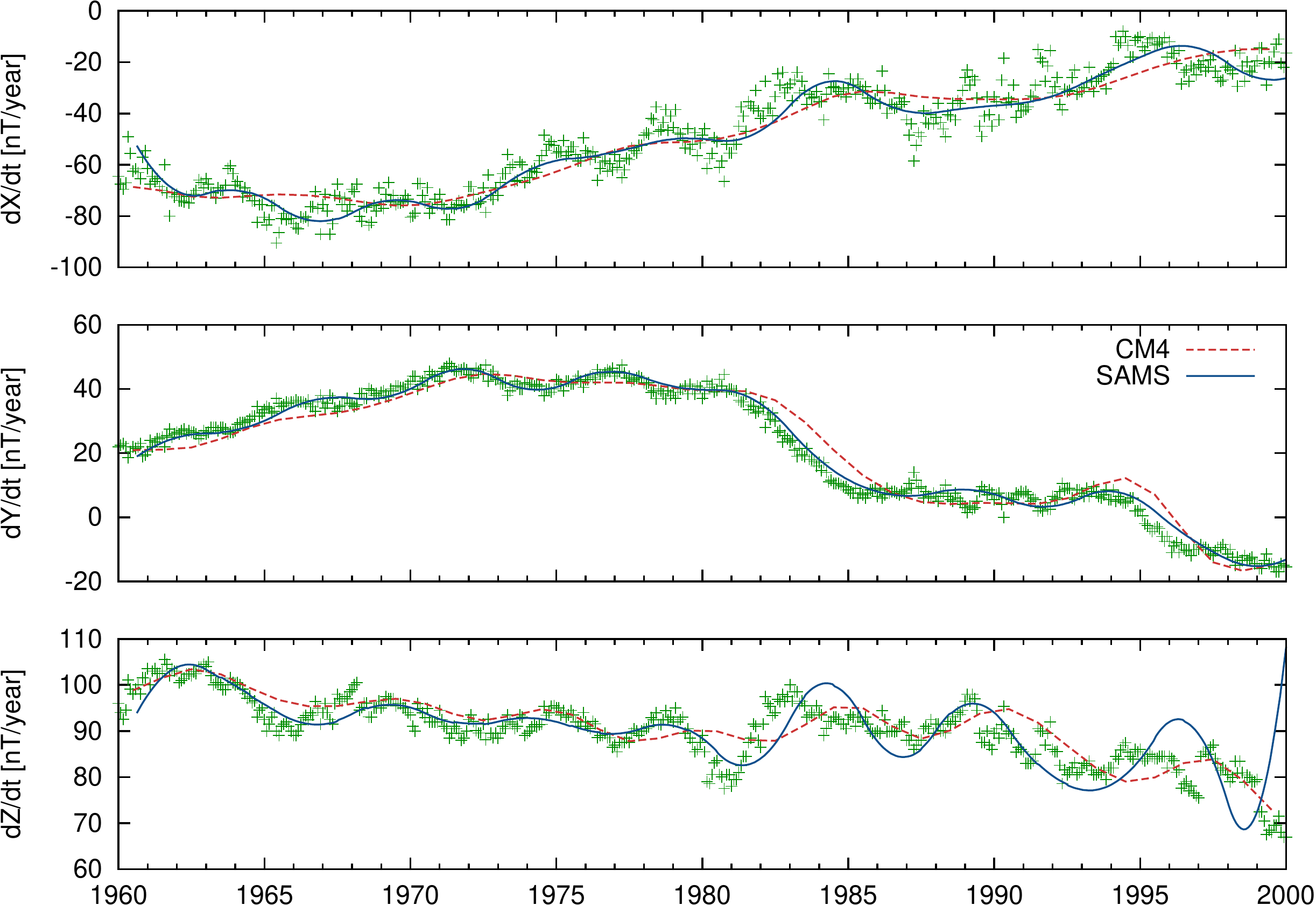}
 \hspace{1cm}
 \includegraphics[width=7cm]{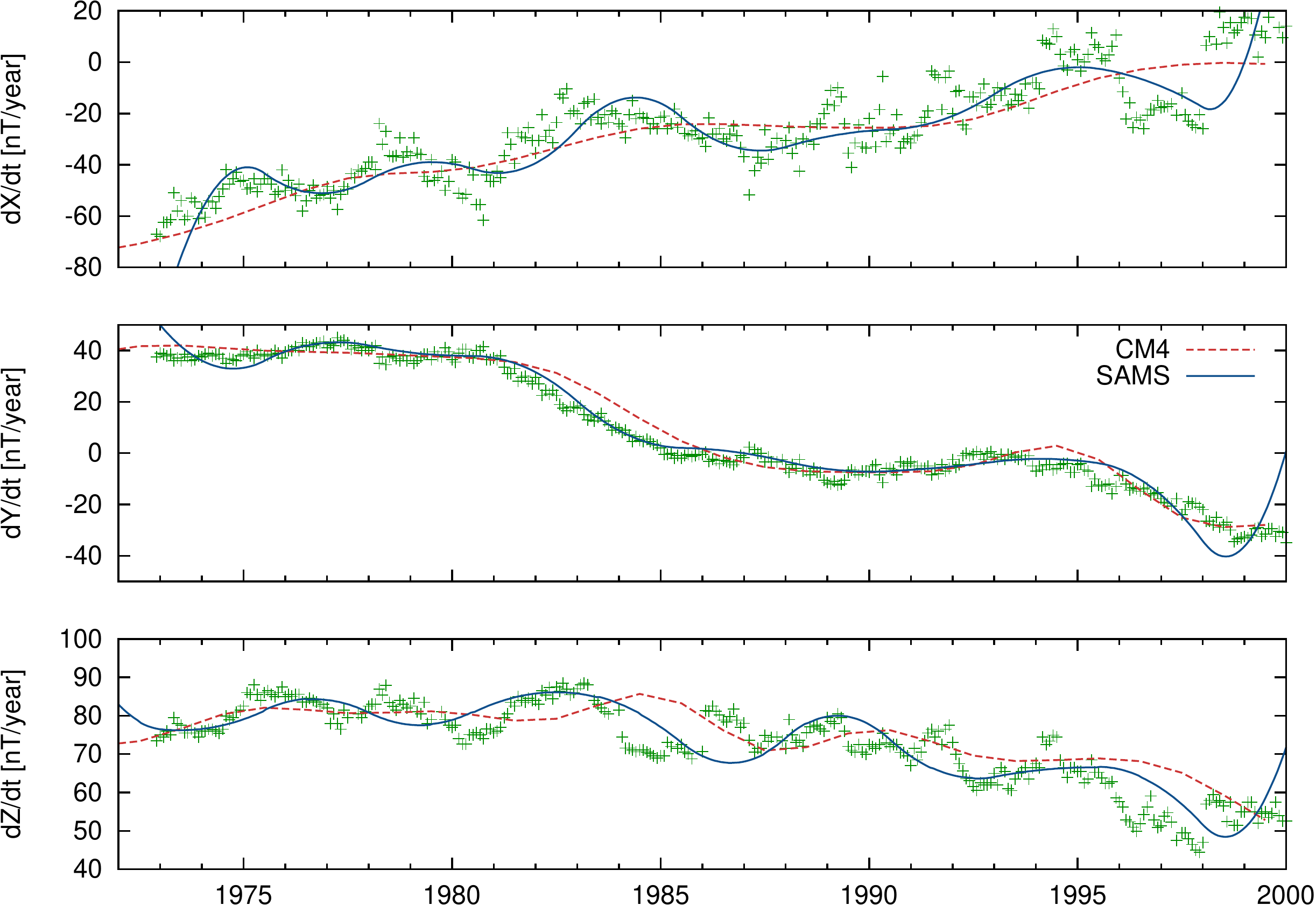}
 \caption{Secular variation at Hermanus and Hartebeesthoek observatories derived from monthly means (crosses) and predicted by SAMS (blue) and CM4 (red) models.}
\label{SV_HER}
\end{figure*}
Probably the most significant improvement of the presented model, compared to previous geomagnetic models over this region, is its ability to map small-scale features of the secular variation. This is illustrated in figure \ref{SV_HER}, which presents observatory monthly mean and modeled values  (from SAMS and CM4) of the first time derivative of the $X$, $Y$, and $Z$ components at Hermanus and Hartebeesthoek observatories. The SAMS model describes well the secular variation in $X$ and $Y$ components. For $Z$ component, the fit is less good, and especially for HBK, temporal edge effects can be observed. However, our model is able to follow more closely the real data than the CM4 model.  We therefore regard SAMS as a suitable model to describe  geomagnetic field changes over this region.\\
\begin{figure*}
 \includegraphics[width=12cm]{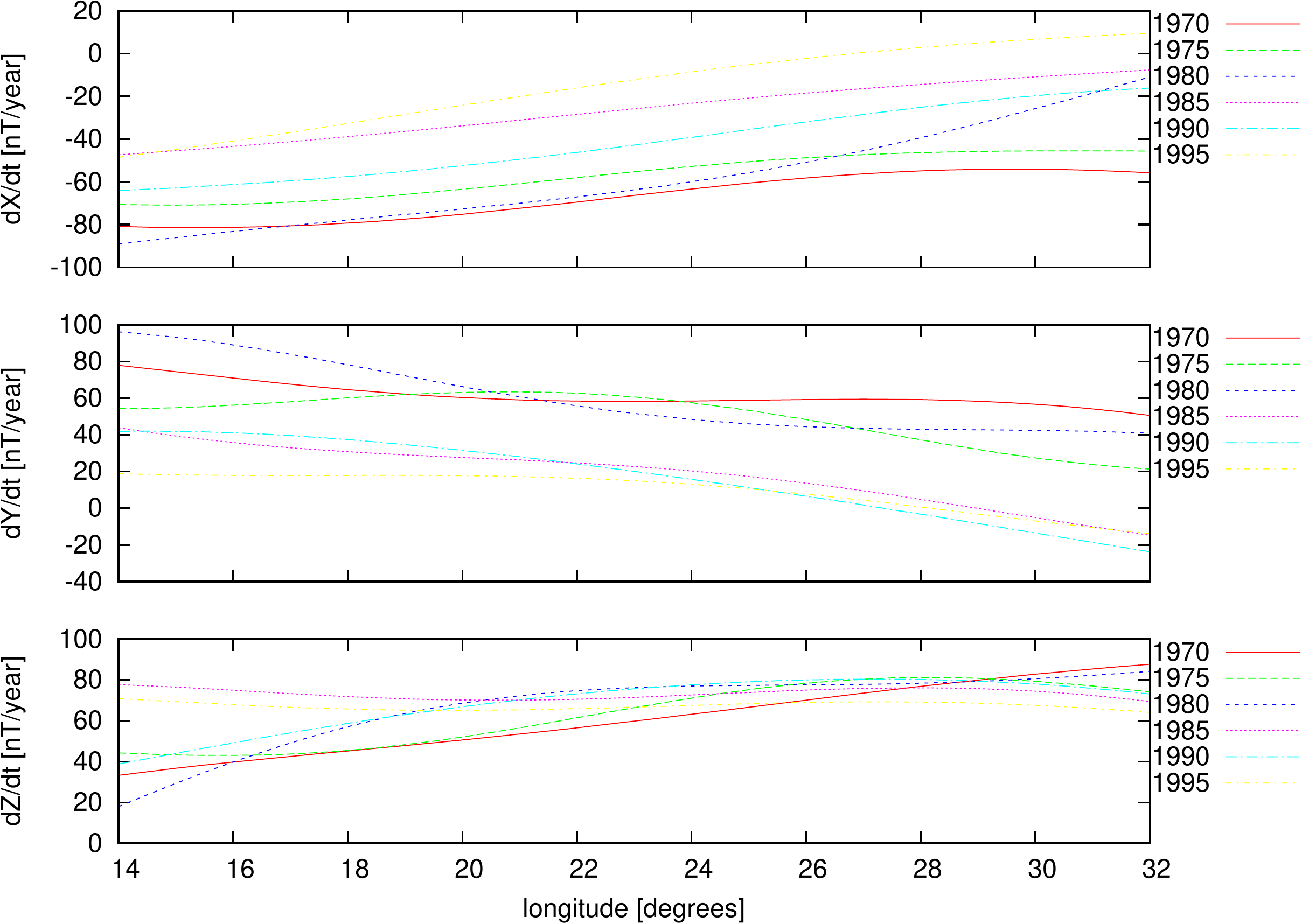}
 \caption{Secular variation estimated from the SAMS model on a profile going from 14$^\circ$ to 30$^\circ$ E longitude at fixed latitude ($\vartheta=25^\circ$ S).}
\label{profile}
\end{figure*}
Although a detailed interpretation of the SAMS  core field and its secular variation is beyond the scope of this paper, we want to point out some interesting behaviour of the secular variation in this region. For this we investigate the changes of the secular variation on different profiles, the one shown here being at latitude 25$^\circ$S, from 14$^\circ$ to 30$^\circ$ E longitude (figure \ref{profile}). The $X$ component has a nearly flat longitudinal gradient, its changes in time being marked by a decrease of $30-50 \mbox{ nT/year}$ from 1975 to 1995. More complicated is the $Y$ component behaviour, for which the changes are dramatically important in the most western part and at the beginning of the period. Finally, remarkable accelerations (from $80 \mbox{ nT/yr}$ to $20 \mbox{ nT/yr}$) are seen in the $Z$ component in the western part, while the secular variation remaining at around the same order of about $80 \mbox{ nT/yr}$ in the eastern part. This value is comparatively high compared to the global rms value of $dZ/dt \sim 66 \mbox{ nT/yr}$ calculated from the IGRF for 1970, which underlines once more the exceptional behaviour of the magnetic field in this region.
\section{Conclusions}
Using observatory and repeat station data, over the southern African region, we have derived a new model, called SAMS, of the internal field over this region. The novelty of this new regional model is the use of harmonic splines. The SAMS brings a more accurate description of the field variation, in space and time, than the available models over this region.\\
In the present study, data over the time interval 1961 - 2001 have been used in order to have a model comparable to CM4 over these four decades. We now plan to extend this initial study, firstly, by taking advantage of the recent (since 2005) efforts to measure the magnetic field on an annual basis. Those data are also available in an unreduced form so that the true date of the measurement could be used. In an area like southern Africa where strong field gradients exist and where observatories are very far apart from each other, the commonly used reduction scheme may lead to increased noise levels. Using the original data, we can avoid the error introduced by reduction to one epoch. A second interesting approach is to consider CHAMP satellite data, and to build a model based on a data set with better spatial coverage and higher temporal resolution. Combining ground-based with satellite data is possible when using harmonic splines for regional modelling, and is one of the next plans for this region.\\
Our easy-to-handle tool for regional modelling is an alternative to the existing regional modelling approaches, which can be used for modelling the Earth's core field and its variations. Finally, let us note that this approach could be included into a global field  model. With the harmonic spline functions, we are able to fit data on places where global models fail and, moreover, to follow fast changes on a regional scale.

 \begin{acknowledgments}
 A.G. was financially partly supported by the Deutsches Zentrum f\"ur Luft- und Raumfahrt under grant 50QW0602.
 We appreciate the work done by our colleagues in Hermanus Magnetic Observatory, in charge of the repeat station network and the magnetic observatories. We would like to express our thanks to Susan Macmillan and Kathryn Whaler who acted as referees and Cor Langereis as editor. All of them gave very constructive comments that improved the manuscript. Maps were produced using the GMT package (\citet{GMT}). This is IPGP contribution XXXX.
\end{acknowledgments}
\bibliographystyle{gji}
\bibliography{lit}
\appendix

\section{Derivation of additional functions} \label{a:add_kernels}
In this appendix, we derive the two additional kernels $F^{L\vartheta}_i$ and $F^{L\varphi}_i$. This is done analogously to the derivation of $F^{Lr}_i$, but considering the two tangential components, $B_\vartheta$ and $B_\varphi$ respectively, instead of the radial component in eq. \ref{B_r}. From this, twe obtain expressions for $F^{L\vartheta}_i$ and $F^{L\varphi}_i$ comparable to the one for  $F^{Lr}_i$ (equation \ref{eq:3.5}).\\
We first express $B_\vartheta$ at a certain point $(\vartheta_i, \varphi_i, r_i)$ 
\begin{eqnarray*} 
\nonumber{B_\vartheta(\vartheta_i, \varphi_i, r_i)}_{}&=&-\frac{1}{r}\frac{\partial V}{\partial \vartheta}\\
\nonumber&=&-\sum^{L}_{l,m}g^m_l \left(\frac{a}{r_i}\right)^{l+2} \partial_\vartheta \left. Y^m_l(\vartheta, \varphi)\right|_{\vartheta=\vartheta_i,\varphi=\varphi_i}\\ 
\nonumber&=&-\sum^{L}_{l,m}4\pi\frac{(l+1)^4l^2}{2l+1} \underbrace{\left[\frac{2l+1}{4\pi(l+1)^4l^2}\left(\frac{a}{r_i}\right)^{l+2}\partial_\vartheta \left. Y^m_l(\vartheta, \varphi)\right|_{\vartheta=\vartheta_i,\varphi=\varphi_i} \right]}_{h^m_{li}}g^m_l\\ 
\nonumber&=&\sum^{L}_{l,m}4\pi\frac{(l+1)^4l^2}{2l+1}\;h^m_{li}\; g^m_l\\ 
&=&<\mathbf{h_i}\;,\;\mathbf{g}>
\end{eqnarray*}
where $\mathbf{h_i}=[\;h^m_{li}\;]_{\{l,m\}}$, $\mathbf{g}=[\;g^m_{l}\;]_{\{l,m\}}$ and  $f_l=\frac{2l+1}{4\pi(l+1)^4l^2}$. The $\mathbf{h_i}$ are linearly independent vectors, and therefore, $\mathbf{g}$ can be written as a linear combination of these vectors:
\begin{equation}
\mathbf{g}=\sum_{j=1}^{N} \alpha_j^\vartheta \mathbf{h_j} .
\end{equation}
Once the $\mathbf{\alpha^\vartheta}$ vector is known, the magnetic field can be computed at any point $(\vartheta, \varphi, r)$ by: 
 \begin{eqnarray} 
\nonumber\mathbf{B}(\vartheta, \varphi, r) &=& -\nabla a \sum^L_{l,m}g^m_l\left(\frac{a}{r}\right)^{l+1}Y^m_l(\vartheta, \varphi)\\ 
\nonumber&=& -\nabla a \sum^L_{l,m} \left\{\sum_j h^m_{lj}\alpha_j^\vartheta\right\}\left(\frac{a}{r}\right)^{l+1}Y^m_l(\vartheta, \varphi)\\
\nonumber&=& -\nabla \sum_j \alpha_j^\vartheta \left\{a\sum^L_{l,m} -\frac{2l+1}{4\pi (l+1)^4l^2} \left(\frac{a}{r_j}\right)^{l+2} \partial_\vartheta \left. Y^m_l(\vartheta, \varphi)\right|_{\vartheta=\vartheta_j,\varphi=\varphi_j} \left(\frac{a}{r}\right)^{l+1} Y^m_l(\vartheta, \varphi)\right\}\\ 
&=&-\sum_j\alpha_j^\vartheta \;\;\nabla F^{L\vartheta}_j(\vartheta, \varphi, r)
\end {eqnarray}
 where we substitute
 \begin{equation} 
 F^{L\vartheta}_i(\vartheta, \varphi, r)=a\sum^L_{l,m} f_l \left(\frac{a}{r_i}\right)^{l+2}\partial_\vartheta Y^m_l(\vartheta_i, \varphi_i) \left(\frac{a}{r}\right)^{l+1} Y^m_l(\vartheta, \varphi)\\
 \end{equation}
Notice that the $f_l$ is the same as in eq. \ref{eq:3.5}.\\
Finally, we consider the $B_\varphi$ component:
\begin{eqnarray} 
\nonumber{B_\varphi(\vartheta_i, \varphi_i, r_i)}_{}&=&-\frac{1}{r \sin \vartheta}\frac{\partial V}{\partial \varphi}\\
\nonumber&=&-\sum^{L}_{l,m}g^m_l \left(\frac{a}{r_i}\right)^{l+2} \left.\frac{\partial_\varphi}{\sin \vartheta}  Y^m_l(\vartheta, \varphi)\right|_{\vartheta=\vartheta_i,\varphi=\varphi_i}\\ 
\nonumber&=&-\sum^{L}_{l,m}4\pi\frac{(l+1)^4l^2}{2l+1} \underbrace{\left[f_l\left(\frac{a}{r_i}\right)^{l+2} \left.\frac{\partial_\varphi}{\sin \vartheta} Y^m_l(\vartheta, \varphi)\right|_{\vartheta=\vartheta_i,\varphi=\varphi_i} \right]}_{s^m_{li}}g^m_l\\ 
\nonumber&=&\sum^{L}_{l,m}4\pi\frac{(l+1)^4l^2}{2l+1}\;s^m_{li}\; g^m_l\\ 
&=&<\mathbf{s_i}\;,\;\mathbf{g}>
\end{eqnarray}
Again, we can write $\mathbf{g}$ as a linear combination, now of the vectors $\mathbf{s_i}$
\begin{equation}
\mathbf{g}=\sum_{j=1}^{N} \alpha_j^\vartheta \mathbf{s_j} ,
\end{equation}
and we rewrite the magnetic field as 
\begin{eqnarray} 
\nonumber B(\vartheta, \varphi, r) &=& -\nabla a \sum^L_{l,m}g^m_l\left(\frac{a}{r}\right)^{l+1}Y^m_l(\vartheta, \varphi)\\ 
\nonumber &=& -\nabla a \sum^L_{l,m} \left\{\sum_j s^m_{lj}\alpha_j^\varphi\right\}\left(\frac{a}{r}\right)^{l+1}Y^m_l(\vartheta, \varphi)\\
\nonumber &=& -\nabla \sum_j \alpha_j \left\{a\sum^L_{l,m} -f_l \left(\frac{a}{r_j}\right)^{l+2}\left.\frac{\partial_\varphi}{\sin \vartheta} Y^m_l(\vartheta,\varphi)\right|_{\vartheta=\vartheta_j,\varphi=\varphi_j} \right.\\
\nonumber &&\hspace{3cm}\times \;\;\;\left. \left(\frac{a}{r}\right)^{l+1} Y^m_l(\vartheta, \varphi)\right\}\\
&=&-\sum_j\alpha_j \;\;\nabla F^{L\vartheta}_j(\vartheta, \varphi, r)
\end {eqnarray}
so that we obtain  
\begin{equation} 
F^{L\varphi}_i(\vartheta, \varphi, r)=a\sum^L_{l,m} f_l \left(\frac{a}{r_i}\right)^{l+2}\frac{\partial_\varphi}{\sin \vartheta} Y^m_l(\vartheta_i, \varphi_i) \left(\frac{a}{r}\right)^{l+1} Y^m_l(\vartheta, \varphi)
\end{equation}

\section{Integrals} \label{a:int}
For the $B_r^r$ component, the integral from equation \ref{int} can be written as
\begin{eqnarray} 
\int_\Omega (\ddot B_r^r)^2 d\omega&=&\int  \sum_{l,m}\sum_{l',m'} \ddot\alpha_i^r \ddot\alpha_j^r (l+1)^2 (l'+1)^2 f_l f_{l'}  \left(\frac{a}{r}\frac{a}{r_i}\right)^{l+2} \left(\frac{a}{r}\frac{a}{r_j}\right)^{l'+2} \cdot \underbrace{Y^{m}_{l}(\vartheta_i, \varphi_i) Y^{m'}_{l'}(\vartheta_j, \varphi_j)}_{P_l^0(\cos\gamma)}  \underbrace{Y^{m}_{l}(\vartheta, \varphi) Y^{m'}_{l'}(\vartheta, \varphi)}_{\frac{4\pi}{2l+1}\delta_{mm'} \delta_{ll'}} d\omega\\
&=&\sum_{i,j} \sum_{l}\ddot\alpha_i^r \ddot\alpha_j^r \frac{4\pi(l+1)^4}{2l+1} f_l^2 \left(\frac{a}{r}\right)^{2l+4} \left(\frac{a}{r_i}\frac{a}{r_j}\right)^{l+2}P_l^0(\cos\gamma)
\end{eqnarray}
using the spherical harmonic addition theorem
\[
 \sum_m Y^{m}_{l}(\vartheta_i, \varphi_i) Y^{m}_{l}(\vartheta_j, \varphi_j)=P_l(cos\gamma)
\]
where $\gamma$ is the angle between $(\vartheta_i, \varphi_i)$ and $(\vartheta_j, \varphi_j)$. $a$ is the reference radius ($6371.2\mbox{ km}$) as introduced in section \ref{method}.\\
The remaining five contributions are more complex
\begin{equation} 
\int_\Omega (\ddot B_r^\vartheta)^2 d\omega=\sum_{i,j} \sum_{l,m}\left[\ddot \alpha_i^\vartheta \ddot \alpha_j^\vartheta \frac{4\pi}{2l+1} (l+1)^2 f_l^2 \left(\frac{a}{r}\right)^{2l+4}  
\times \left(\frac{a}{r_i}\frac{a}{r_j}\right)^{l+2} \partial_\vartheta Y^m_l(\vartheta_i, \varphi_i)\partial_\vartheta Y^m_l(\vartheta_j, \varphi_j)\right]
\end{equation}
\begin{equation} 
\int_\Omega (\ddot B_r^\varphi)^2 d\omega=\sum_{i,j} \sum_{l,m}\left[\frac{\ddot \alpha_i^\vartheta \ddot \alpha_j^\vartheta}{\sin\vartheta_i \sin\vartheta_j} \frac{4\pi}{2l+1} (l+1)^2 f_l^2 \left(\frac{a}{r}\right)^{2l+4}\times\left(\frac{a}{r_i}\frac{a}{r_j}\right)^{l+2} \partial_\varphi Y^m_l(\vartheta_i, \varphi_i)\partial_\varphi Y^m_l(\vartheta_j, \varphi_j)\right]
\end{equation}
\begin{equation} 
\int_\Omega\ddot  B_r^r \ddot B_r^\vartheta  d\omega=\sum_{i,j} \sum_{l,m}\left[\ddot \alpha_i^r \ddot \alpha_j^\vartheta \frac{4\pi}{2l+1} (l+1)^3 f_l^2 \left(\frac{a}{r}\right)^{2l+4} \times \left(\frac{a}{r_i}\frac{a}{r_j}\right)^{l+2}  Y^m_l(\vartheta_i, \varphi_i)\partial_\vartheta Y^m_l(\vartheta_j, \varphi_j)\right]
\end{equation}
\begin{equation} 
\int_\Omega \ddot B_r^r \ddot B_r^\varphi d\omega=\sum_{i,j} \sum_{l,m}\left[\frac{\ddot \alpha_i^r \ddot \alpha_j^\vartheta}{ \sin\vartheta_j} \frac{4\pi}{2l+1} (l+1)^3 f_l^2 \left(\frac{a}{r}\right)^{2l+4} \times \left(\frac{a}{r_i}\frac{a}{r_j}\right)^{l+2}  Y^m_l(\vartheta_i, \varphi_i)\partial_\varphi Y^m_l(\vartheta_j, \varphi_j)\right]
\end{equation}
\begin{equation} 
\int_\Omega \ddot B_r^\vartheta \ddot B_r^\varphi d\omega=\sum_{i,j} \sum_{l,m}\left[\frac{\ddot \alpha_i^\vartheta \ddot \alpha_j^\varphi}{\sin\vartheta_j} \frac{4\pi}{2l+1} (l+1)^2 f_l^2 \left(\frac{a}{r}\right)^{2l+4} \times  \left(\frac{a}{r_i}\frac{a}{r_j}\right)^{l+2} \partial_\vartheta Y^m_l(\vartheta_i, \varphi_i)\partial_\varphi  Y^m_l(\vartheta_j, \varphi_j)\right]
\end{equation}

\label{lastpage}
\end{document}